# Phloem loading through plasmodesmata: a biophysical analysis.

Jean Comtet[†,*], Robert Turgeon[‡,*] and Abraham D. Stroock[†,§,*]

[†]School of Chemical and Biomolecular Engineering,[‡] Section of Plant Biology, and [§]Kavli Institute at Cornell for Nanoscale Science, Cornell University, Ithaca, NY 14853.

*Correspondence: Jean Comtet (jean.comtet@gmail.com), Robert Turgeon (ert2@cornell.edu) and Abraham D. Stroock (ads10@cornell.edu)

**Abstract**

In many species, sucrose en route out of the leaf migrates from photosynthetically active mesophyll cells into the phloem down its concentration gradient via plasmodesmata, i.e., symplastically. In some of these plants the process is entirely passive, but in others phloem sucrose is actively converted into larger sugars, raffinose and stachyose, and segregated (trapped), thus raising total phloem sugar concentration to a level higher than in the mesophyll. Questions remain regarding the mechanisms and selective advantages conferred by both of these symplastic loading processes. Here we present an integrated model – including local and global transport and the kinetics of oligomerization – for passive and active symplastic loading. We also propose a physical model of transport through the plasmodesmata. With these models, we predict that: 1) relative to passive loading, oligomerization of sucrose in the phloem, even in the absence of segregation, lowers the sugar content in the leaf required to achieve a given export rate and accelerates export for a given concentration of sucrose in the mesophyll; and 2) segregation of oligomers and the inverted gradient of total sugar content can be achieved for physiologically reasonable parameter values, but even higher export rates can be accessed in scenarios in which polymers are allowed to diffuse back into the mesophyll. We discuss these predictions in relation to further studies aimed at the clarification of loading mechanisms, fitness of active and passive symplastic loading, and potential targets for engineering improved rates of export.

Abbreviations: *M, Mesophyll; P, Phloem; RFOs, Raffinose Family Oligosaccharides; MV, Minor Vein*

Vascular plants export sugars and other nutrients from leaves through a living vascular tissue, the phloem. Export distributes photosynthetic products to remote tissues (sinks) for growth and storage and couples synthesis and intercellular transport processes in the leaves and sink tissues to global, hydraulic transport through the phloem sieve tubes and xylem vessels. Significant uncertainties remain regarding the structure, chemistry and transport phenomena governing these processes (1–3). Improved models of export will inform our understanding of whole-plant physiology and open opportunities to engineer sugar concentrations and transport processes to improve growth and yield (4, 5). Insights into these transport processes may also suggest ways to design efficient synthetic systems to control chemical processes (6, 7).

Particular outstanding questions relate to the mechanisms by which plants transfer sucrose, and in some cases sugar alcohols, from the photosynthetically active mesophyll to the transport phloem (phloem loading) in the sub-set of species in which this loading step occurs symplastically, i.e., through the open channels of plasmodesmata (8, 9) (Fig. 1). In most symplastic loaders there is no buildup of sugar in the phloem, as shown in Fig. 1*B*; this distribution of sugars suggests passive transfer from mesophyll to phloem, as postulated by Münch (10). In a second symplastic loading mechanism, sucrose passes into the phloem through specialized plasmodesmata (Fig. 1*A*) and is converted, in an energetically active process, to raffinose family oligosaccharides (RFOs, principally raffinose and stachyose) in the phloem companion cells. Transfer of RFOs back into the mesophyll does not appear to occur and one observes elevated total concentrations of sugars in the phloem relative to the mesophyll (Fig. 1*B*) (11–13). This inversion of the concentration gradient depends on the polymerization reaction (14) and correlates with lower sugar concentration in the mesophyll, and hence in the whole leaf relative to plants that load passively (15) (Fig. 1*C*). In these two characteristics (inverted concentration gradient and lower total sugar content in leaves), active symplastic loaders, also known as polymer trappers, match the behavior of apoplastic loaders in which photoassimilate is actively pumped into the phloem (Fig. 1C) (11, 12, 15). In this paper, we refer to RFO accumulation in the phloem as "segregation" and the elevated total sugar concentration in the phloem relative to the mesophyll as "gradient inversion".

The observation of strong sugar segregation in the phloem (Fig. 1*B*) and low levels of whole leaf sucrose (Fig. 1*C*) in polymer trap plants provokes a number of questions. First, what mechanisms permit passive transport of sucrose through these apparently open pores from mesophyll to minor vein phloem, while simultaneously preventing the passage of larger RFOs in the opposite direction? One possibility is that the plasmodesmata in question are very narrow, allowing sucrose to pass via diffusion (16–18) or convection (11, 19) while inhibiting RFO backflow on the basis of steric selectivity. However, coupling of local plasmodesmatal dynamics with whole plant transport of water and sugar and the kinetics of polymerization has so far been neglected. A second question is raised by segregation: how can phloem osmolarity be higher than in the mesophyll given that oligomerization reactions reduce the number of osmotically active molecules in the phloem sap? Finally, a more general question: how do the rates of symplastic loading, convective export and polymerization influence sugar segregation and translocation rates?

Only a few models of phloem transport consider loading mechanisms and distinguish between mesophyll and phloem (19–21). Other simplified modeling approaches (22–24) have given insight into phloem traits at the plant scale but avoid the question of phloem loading by considering a fixed hydrostatic pressure in the phloem. In this paper, we introduce a global model of water and sugar transport in symplastic loading species with explicit kinetics of polymerization (Fig. 1*D-E*). We then consider the transport properties of plasmodesmata, including the relative importance of diffusion and convection and determine how long-distance transport is affected by segregating sucrose polymers in the phloem. These analyses provide new insights into the nature of symplastic loading mechanisms and the adaptive advantages they confer.

**A globally coupled model of symplastic loading with polymerization**

Fig. 1*D* is a schematic cross-section of a leaf minor vein in a symplastic loader (electron micrograph in Fig. S2) presenting the hypothesized transport processes: Photosynthetic products (red circles; sucrose in all plants and sugar alcohols in some) diffuse and convect through plasmodesmata (cross-sectional view in Fig. 1*A*) down their concentration gradient from the mesophyll (site of synthesis) to the phloem (site of convective evacuation). Sucrose is then polymerized into RFOs (green double circles in Fig. 1*D*). Elevated osmolarity in the mesophyll and phloem recruits water from the xylem (blue arrows) to drive convection along this pathway. Water and sugars are subsequently exported by convection through the transport phloem (T-Phloem) to sinks (blue and red downward arrows, respectively).

Fig. 1*E* presents a circuit representation of steady fluxes of water (blue arrows) and sugar (red arrows) from xylem to the mesophyll ($Q_{XM}$ [m s$^{-1}$]) and to the MV-phloem ($Q_{XP}$), from mesophyll to the MV-phloem ($Q_{MP}$; $\phi_{MP}^{suc}$ [mole m$^{-2}$ s]; $\phi_{MP}^{RFO}$), and through the phloem to sink tissues ($Q_P$; $\phi_P^{suc}$; $\phi_P^{RFO}$). All fluxes are defined with respect to the exchange surface area of MV-phloem through which sucrose loading occurs. The zig-zag black lines represent paths for water and solute transfer. Each path presents a hydraulic conductance ($L$ [m s$^{-1}$ Pa$^{-1}$]) for water flow. The interface with the xylem is a perfect osmotic membrane that excludes solute passage by either convection (reflection coefficient, $\sigma_{XM} = \sigma_{XP} = 1$) or diffusion (diffusive mass transfer coefficient, $k_{XM} = k_{XP} = 0$ (m s$^{-1}$)) (25). The plasmodesmatal interface between the mesophyll and phloem partially reflects solute ($0 \leq \sigma_{MP} \leq 1$) and allows for diffusive solute transfer ($k_{MP} \geq 0$); we explore details of plasmodesmatal transport processes in Fig. 2. The transport phloem allows water flow and free convective solute transfer ($\sigma_P = 0$) neglecting diffusion ($k_P = 0$). We consider Michaelis-Menten kinetics for polymerization of *n* sucrose into one RFO, with a maximal rate, $\phi_{pol}^{MM}$ [mole m$^{-2}$ s]. With $\phi_{pol}^{MM} = 0$, the system models passive symplastic loading. See *Materials and Methods* and *SI Text S1* for details.

***Non-dimensional parameters that characterize loading.*** Coupled convection and diffusion define the pressures, concentrations, and fluxes in the loading zone (Fig. 1*E*). Before proceeding, we identify generic features of this coupling. First, we identify the characteristic net driving force for water flow from leaf to sink $\Delta P_c$ as the typical mesophyll osmotic pressure minus the adverse pressure difference between leaf xylem and the unloading zone in the transport phloem:

$$\Delta P_c = RT c_M^{suc,0} + P_X - P_S, \qquad (1)$$

Second, analysis of the hydraulic network gives a total conductance for the leaf in series with the transport phloem:

$$L_{tot} = \frac{1}{\frac{1}{L_{leaf}} + \frac{1}{L_P}}, \qquad (2)$$

where $L_{leaf} = 1/(1/L_{XM} + 1/L_{MP}) + L_{XP}$ is the effective conductance of the leaf ($L_{XM}$ is in parallel with $L_{XM}$ and $L_{MP}$, which are in series). Together, Eqs. 1 and 2 define water flux through the phloem:

$$Q_P^c = L_{tot} \Delta P_c = L_{tot}[RT c_M^{suc} + P_X - P_S], \qquad (3)$$

This flow carries sugars out of the MV-phloem at a rate, $\phi_P = c_M^{suc} Q_c$, so that we expect that the sugar concentration in the phloem will depend, in part, on a competition between this convective transfer and sucrose diffusion through the plasmodesmata interface. To characterize this competition, we propose the following non-dimensional ratio of global convection and local diffusion:

$$f \equiv \frac{\text{convection}}{\text{diffusion}} = \frac{Q_P^c c_M^{suc}}{k_{MP}^{suc} c_M^{suc}} = \frac{Q_P^c}{k_{MP}^{suc}} = \frac{L_{tot}[RT c_M^{suc} + P_X - P_S]}{k_{MP}^{suc}}, \quad (4)$$

where $k_{MP}^{suc}$ [m s$^{-1}$] is the diffusive mass transfer coefficient through the plasmodesmatal interface. For large values of this flushing number, $f$, phloem loading is diffusion-limited and the concentration of phloem sugars will be low because solutes are flushed out of the MV-phloem more quickly that they can diffuse in; gradient inversion (elevated total concentration of sugars in MV-phloem – Fig. 1B-C) is suppressed in this regime. For small values of $f$, loading is convection-limited and sugar concentration in the MV-phloem is high, favoring gradient inversion. This number is relevant for both passive and active symplastic loaders (7).

**Physiology of the plasmodesmata**.

Although segregation of RFOs based on a size exclusion mechanism has been proposed (19, 26), discrimination based only on hydrodynamic radii seems difficult considering that stachyose is only 40% larger than sucrose (17) and raffinose is even smaller. Even if the stachyose mass transfer coefficient is reduced by steric interaction with plasmodesmatal channels (18, 19), at steady state back diffusion of raffinose and stachyose into the mesophyll will eventually occur. Dölger et al. (19) presented a phenomenological model of hindered transport of water and sucrose through the plasmodesmata interface, suggesting that back-flux could be prevented by water flow, concluding that this mechanism is not feasible. Here, we present an explicit model of convection and diffusion within the plasmodesmata (Fig. 2) to use with our global model and reexamine the mechanism of RFO segregation (Fig. 3).

*Pore-scale model of plasmodesmata transport.* Fig. 2*A* presents an electron micrograph of a plasmodesma in transverse section. Sugar molecules are thought to pass through the space between the desmotubule (Fig. 2*A*, "DW") and the plasma membrane (Fig. 2*A*, "IPM"), the "cytoplasmic sleeve." One idealized interpretation the cytoplasmic sleeve is as a series of nanochannels created by regularly arranged proteins (Fig. 2*A*, "S") (27, 28). In Fig. 2*B*, we model each plasmodesma as a bundle of 9 pores of equivalent radius $r_{pore}$ and length $L$ (28).

In Fig. 2C, we follow Deen (29) in considering hindered transport for spherical solutes in cylindrical pores with purely steric interactions. We introduce a confinement parameter, the ratio of pore radius to solute radius, $\lambda_i = r_i/r_{\text{pore}}$; this parameter controls the partial rejection of solute species $i$ due to steric interactions with the pore, such that the sugar flux from mesophyll to phloem can be expressed as:

$$\phi_{\text{MP}}^i = [1 - \sigma_{\text{MP}}^i(\lambda_i)]Q_{\text{MP}} \exp\left[c_M^i + \frac{c_M^i - c_P^i}{\exp(\text{Pe}_i) - 1}\right], \quad (5)$$

where $\sigma_{\text{MP}}^i(\lambda_i)$ is the reflection coefficient that depends only on the ratio $\lambda_i$, $\text{Pe}_i = [1 - \sigma_{\text{MP}}^i(\lambda_i)]Q_{\text{MP}}/k_{MP}^i$ is the ratio of hindered convection to hindered diffusion in the pore, and $k_{MP}^i$ is the mass transfer coefficient that accounts for hindered diffusion (29, 30) (*Eq. S18-21*). In the limit of $\text{Pe}_i \ll 1$, Eq. (5) simplifies to $\phi_{\text{MP}}^i = k_{MP}^i(c_M^i - c_P^i)$, corresponding to purely diffusive flux through plasmodesmata.

***Together, steric hindrance and convection can inhibit RFO transfer from phloem to mesophyll.*** In Fig. 2*D*, we plot solute flux through such a model pore as a function of the confinement parameter $\lambda$. We normalize total flux, $\phi_{\text{MP}}$ by its diffusive component, $k_{\text{MP}}|\Delta c| > 0$. Positive flux corresponds to net transfer from mesophyll (left) to phloem (right). Flux is driven by a fixed water potential difference, $\Delta\psi_{\text{MP}} = \Delta P - \sigma RT\Delta c = 0.1$ MPa, with $\psi_M > \psi_C$ and fixed solute concentrations at the ends of the pore, with the sucrose concentration (red) high in the mesophyll and the RFO concentration (green) high in the phloem. The upper axis of Fig. 2D represents the reflection coefficient, $\sigma_{\text{MP}}(\lambda)$ (30) for a given confinement parameter (Eq. S21). The pore wall strongly impedes diffusive transport by increasing viscous drag experienced by solute particles, while advection of solute is less hindered, as steric interactions restrict solute to the zone of maximum flow in the center of the pore, where convection is strongest (Fig. 2*C*). Thus, for very narrow pores, convective transport of solute dominates diffusive transport and upstream transfer of RFOs (from phloem to mesophyll) is suppressed as $\lambda \rightarrow 1$ (Fig. 2*D*, green solid line above 0 for small $\lambda$). For larger pores ($\lambda > 5$ here), we also predict segregation of RFO, because convection dominates again in this limit.

In summary, we predict that convection inhibits back diffusion of RFO from phloem to mesophyll in the limits of both strong ($\lambda \rightarrow 1$) and weak ($\lambda \gg 1$) confinement within the plasmodesmata. This prediction, with a more complete treatment of plasmodesmatal transport, contradicts the conclusion of Dölger et al. (19) that flow cannot prevent back diffusion of RFO and opens this alternative route to segregation and gradient inversion in symplastic loaders.

**Whole plant transport and plasmodesmatal selectivity**.

Fig. 3 presents predictions of our global water transport model (Figs. 1*D*-1*E*) with the hindered transport model presented in Fig. 2. Figs. 3*A-B* present sugar distributions (3*A*) and sucrose export rate (3*B*) with respect to the strength of global convection versus diffusion (*f*) and relative pore size ($\lambda_{\text{RFO}} = r_{\text{pore}}/r_{\text{RFO}}$ on left axis; $\lambda_{\text{suc}} = r_{\text{pore}}/r_{\text{suc}}$ on right axis) for a typical polymerization rate (see *Materials and Methods)*. We use parameters for stachyose to represent RFO species, with degree of polymerization n=2. The charts in Fig. 3*C* present calculated concentrations of RFO (green) and sucrose (red) in the mesophyll cells (M) and phloem (P) at three points of differing convection and hindrance. In Figs. 3D-F, we explore trends with polymerization rate. (*Table S1* for parameter values).

*Gradient inversion can occur without chemically selective plasmodesmata.* The solid red line in Fig. 3*A* ("1:1") represents the boundary between the states that show gradient inversion and those that do not: below this curve, for lower $\lambda_{\text{RFO}}$ and $\lambda_{\text{suc}}$ (more restricted motion within the plasmodesmata) and lower flushing number, the total sugar concentration in the phloem is higher than in the mesophyll. The other curves represent states with excess concentrations of sugar in the phloem relative to the mesophyll. Importantly, for parameters within the physiological range (unshaded area), we predict that inversion can occur, with magnitudes of excess concentration in the phloem (50-100% - point (1) on the diagram) that match those observed in active symplastic loaders (15). This gradient inversion depends on two conditions: i) strong geometric confinement within the plasmodesmata ($\lambda_{\text{RFO}} < 1.3$), for which convection of RFO from mesophyll to phloem tends to overcome its back diffusion, corresponding to the limit of $\lambda \to 1$ on the solid green curve in Fig. 3*D*; and ii) weak convection through the phloem (low *f*). If either of these conditions is violated, gradient inversion fails to occur: with strong hindrance and strong convection (point (2)), segregation of RFO occurs (green bars in Fig. 3*C*, point (2)), but sugars are flushed out of the phloem, prohibiting gradient inversion. With weak hindrance and weak convection (point (3)), segregation of RFO does not occur and the gradient between the mesophyll and the phloem tends to zero (Fig. 3*C*, point (3)).

In Fig. 3D, we plot the ratio of the total concentrations of sugar ($c^{\text{tot}} + c^{\text{RFO}}$) in the phloem and mesophyll for a fixed mesophyll sucrose concentration. We associate values of $c_{\text{P}}^{\text{tot}}/c_{\text{M}}^{\text{tot}} > 1$ with gradient inversion (above the dashed line in Fig. 4F). We see that, for weak global convection (low $f$) and hindered transport through plasmodesmata ($\lambda_{\text{RFO}} \to 1$) (1 – blue curve), the strength of gradient inversion grows monotonically with polymerization rate, confirming that an increase in diffusive flux created by the sucrose depletion in the phloem overcomes unfavorable stoichiometry with respect to total moles of solute. For flushing numbers greater than one (2 – red curve) or weak segregation (3 – yellow curve) gradient inversion is never obtained, even for large polymerization rates.

*Export rates are compatible with those observed experimentally.* Based on the predictions in Fig. 3*A*, the experimentally observed degree of gradient inversion requires strongly hindered transport through the plasmodesmata (small $\lambda_{\text{RFO}}$ as at point (1)). To assess the consequences of this hindrance on sugar flux, in Fig. 3*C* we plot the equivalent sucrose export rate, $\phi_{\text{suc}}^{\text{eq}}$ over the same domain as in 3*A*. The green curve is the isoline for an export rate, $\phi_{\text{suc}}^{\text{eq}} = 0.9$ μmol/m²/s corresponding to a typical flux through minor veins (12, 31). Importantly, the model is consistent with experiments in that strong gradient inversion occurs in a regime that provides physiologically reasonable export rates (as at point (1)). However maintaining gradient inversion significantly constrains export rates compared to those given larger plasmodesmatal pores (as at point (3)). Importantly, this suggests that segregation of RFOs and gradient inversion do not provide a direct advantage with respect to export rates, and that the elevated density of plasmodesmata observed in active symplastic loaders relative to passive ones may have evolved to accommodate the limitation on flux imposed by the narrow pores required for gradient inversion (32).

*Polymerization lowers the required concentration of sugars in the leaf.* We now explore the impact of polymerization on total sugar concentration in the leaf for a fixed rate of synthesis in the mesophyll, $\phi_{\text{syn}}$ or export through the phloem (these rates are equal in equivalent moles of sucrose at steady state). In Fig. 3E we track the average sugar concentration of the entire leaf, $c_{\text{leaf}} = v_{\text{M}} c_{\text{M}}^{\text{suc}} + v_{\text{P}}(c_{\text{P}}^{\text{suc}} + c_{\text{M}}^{\text{RFO}})$ as a function of the rate of polymerization, $\phi_{\text{pol}}^{\text{MM}}$ for the three cases highlighted in Figs. 3A-3C. In the definition of $c_{\text{leaf}}$, $v_{\text{M}}$ and $v_{\text{P}}$ are the volume fractions of mesophyll and phloem in a typical leaf. The values of $c_{\text{leaf}}(f)$ at $\phi_{\text{pol}}^{\text{MM}} = 0$ correspond to passive loading. Importantly, Fig. 3E shows that the total leaf sugar concentration required to drive export always decreases with increasing polymerization rate. To maintain a given sucrose flux, the difference in sucrose concentration must be maintained at a fixed value to drive diffusion; increasing the rate of polymerization lowers the phloem sucrose concentration and allows the concentration in the mesophyll to drop while maintaining a fixed gradient. Due to the stoichiometry of polymerization and the small volume fraction occupied by the phloem ($v_P \ll 1$), the RFOs produced contribute negligibly to total leaf sugar content, even in the absence of segregation (point (3)). This prediction supports the proposal that polymerization provides a selective advantage by lowering sugar content in leaves to increase growth potential and minimize herbivory (3, 15).

***Increased polymerization and convection increase export rate.*** We now ask how polymerization and convection impact export rate with a fixed concentration of sucrose in the mesophyll. We track the equivalent molar flux of sucrose out of leaves,

$$\phi_{\text{suc}}^{\text{eq}} = Q_P\big(c_P^{\text{suc}} + n c_P^{\text{RFO}}\big), \qquad (6)$$

as a function of $\phi_{\text{pol}}^{\text{MM}}$. For a fixed mesophyll sucrose concentration, the equivalent sucrose flux always increases with polymerization rate (Fig. 3F). This effect is due to the increased gradient in sucrose concentration created by sucrose depletion in the phloem by polymerization. Fig. 3F also shows that export rate increases with increased convection (higher $f$, comparing (1) and (2)). We note that the favorable dependence of translocation rate on polymerization holds only if RFO synthesis is spatially confined to the phloem because the reaction must selectively decrease the concentration of sucrose in the phloem to increase the gradient between the two cellular domains; this confinement of the enzymes has been reported for active symplastic loaders (33, 34).

***Oligomerization and segregation minimize sugar content of leaves.*** We now explore the possible advantages derived from the polymer trap phenomenon (Figs. 3*D-F*). Note that increased rate of polymerization lowers $c_{\text{leaf}}$ (Fig. 3*E*) and increases $\phi_{\text{suc}}^{\text{eq}}$ (Fig. 3*F*) regardless of the degree of gradient inversion (3*D*). The notable distinction of the strongly hindered case (point (1) – blue curves) is that it displays both strong gradient inversion and a rapid decay of $c_{\text{leaf}}$ with $\phi_{\text{pol}}^{\text{MM}}$; in leaves operating under these conditions, a small expenditure of metabolic activity dedicated to oligomerization will dramatically decrease its load of sugar. This suggests that maintaining low sugar content in leaves provides a selective advantage for the specialized plasmodesmata that lead to segregation and gradient inversion in active symplastic loaders.

**Discussion**

The physico-chemical mechanism of active symplastic loading has remained obscure, as have the relationships between various of its experimentally observed characteristics and biological functions. To shed light on these topics, and on symplastic loading more generally, we have introduced a model that couples local and global transport processes with the oligomerization kinetics of sucrose into RFOs.

Our predictions indicate that, regardless of global hydraulic conditions, localized oligomerization of sucrose into RFOs in the MV-phloem decreases the total concentration of sugars required in the leaf to export sucrose at a fixed rate (Fig. 3*E*) and increases the rate of export for a fixed concentration of sucrose in the mesophyll (Fig. 3*F*); both of these trends could be beneficial to the plant and provide a basis for a selective pressure toward this metabolically active reaction (3). With the introduction of a simple but complete model of hindered convection and diffusion within the plasmodesmata, we find that the conditions required to provide segregation and gradient inversion lead to physiologically reasonable rates of export, if account is taken for the unusually high density of plasmodesmata in trapper species. While higher export rates could be achieved for conditions that do not provide gradient inversion (larger pore radii and higher flushing number), these conditions do not lead to as large a reduction of sugar concentration in the mesophyll as the strongly segregated case (Fig. 3*E*). Taken together, our observations are consistent with the hypothesis that the specialized plasmodesmata found in active symplastic loaders – with high areal density and nanometer-scale effective pore radii – evolved to provide an adequate export rate (e.g., a value limited by photosynthetic rates) under the additional constraint of minimizing the total sugar content of leaves (Fig. 1*C*) (15). Turgeon argued that reducing total carbohydrate concentration in the leaves could increase growth potential and limit herbivory (3). We also note that minimizing sugar concentration, and in particular RFOs, in the mesophyll could minimize possible inhibition of photosynthesis (35). A clear prediction is that a strong decrease in the convective flow through the plasmodesmata should impede segregation if selectivity is provided by hindered plasmodesmatal transport. Along with dye coupling approaches (36), experiments decreasing the flushing number (by applying cold, or girdling the transport phloem), could give additional insight into RFOs segregation mechanisms.

Interestingly, most, if not all, symplastic loaders oligomerize some sucrose into RFOs whether or not they display the gradient inversion associated with polymer trapping (32). Our model provides a possible rational for this observation: with or without segregation and gradient inversion, localized reduction of sucrose in the phloem by oligomerization increases export rates relative to the completely passive case (as for points (2) and (3) in Fig. 3). While trapping species appear to use segregation to prioritize low concentrations in the mesophyll, other symplastic loaders may be exploiting this effect to a lesser degree, prioritizing export rate over the minimization of concentration. In other words, we suggest that oligomerization may represent an active loading process in a much larger fraction of symplastic loaders than has been previously appreciated. It would be interesting to confirm the predicted relation between polymerization and translocation experimentally by genetically enhancing or inhibiting polymerization rates in both symplastic loaders that show gradient inversion and those that do not (14, 37). Such techniques could potentially play a role in improving phloem export rates and yields in symplastic loaders.

Our model of transport through plasmodesmata shows that both segregation of RFOs in the phloem and gradient inversion can occur without strict steric exclusion or chemical selectivity (Fig. 3*A*). We conclude that convective sweeping of RFOs downstream in the plasmodesmata (from mesophyll to phloem) plays a critical role in driving these effects, in contrast to the conclusions of a recent study (19). We do not exclude the possibility that molecular mechanisms (e.g. due to molecularly specific steric or chemical effects in the pores) could impact the selectivity for transfer of sucrose relative to RFOs. Alternatively, as noted by Liesche and Schulz (17), stachyose diffusing into the mesophyll could be hydrolyzed back into sucrose and monosaccharides by the alpha-galactosidase present in mature leaves, preventing stachyose accumulation in the mesophyll. To clarify this mechanism further will require additional information on sugar gradients and hydraulic coupling between mesophyll and phloem (11), the structure and biochemistry of the pore spaces within plasmodesmata, and more detailed models of molecular transport under strong confinement (38). Our model provides a framework in which to evaluate the impact of these details on global properties of the loading process.

In conclusion, our study highlights the impact of system-scale coupling on the dynamics of symplastic loading and sheds light on the possible selective advantages derived from the polymerization and segregation that are observed in polymer trap species. We propose that that evolutionary drivers other than increased export rate should be sought to explain sugar segregation in active symplastic loaders and that upregulation of the enzymatic pathways that synthesize RFO could lead to improved export rates in passive symplastic loaders. In conjunction with future experiments, refinements of this model could provide a basis for directing the design of engineered plants with more efficient translocation of sugars, faster growth, and higher yields.

## Materials and Methods

**Boundary condition for mesophyll sucrose.** Photosynthesized carbohydrates can selectively be stored as starch or sucrose. This partitioning led us to consider, in Fig. 3, two extreme boundary conditions for the export and concentration of mesophyll sucrose. When photosynthesis is not limiting we consider sucrose concentration to be fixed at 200 mM in the mesophyll (Figs. 3A-D and 3F). For limiting photosynthesis, all sugars are exported and a fixed sugar flux $\phi_{\text{MP}}^{\text{syn}} = 900$ nmol/m2/s has to be accommodated through the phloem (12, 31) (Fig. 3E). (*SI Text*, S1 and Eq. S11 for more details).

**Coupling to the Xylem.** Windt et al. (39) showed experimentally that the impact of phloem flow on xylem water status is weak. We thus take fixed water pressure in the xylem, $P_X = -0.1$ MPa to represent leaves at moderate stress. Change in xylem water potential would simply shift the flushing number $f$. Mesophyll cells and minor veins are surrounded by cell walls, which, as part of the apoplast, can lead to direct hydraulic coupling to the xylem. Due to differences in water potential, water from the xylem can enter these cells via osmosis, (Figs. 1D-E blue arrows) after diffusing through the cell walls and membrane aquaporins (40). In Fig. 3, we take the equivalent water permeability of these interfaces to be $L_{\text{XP}} = L_{\text{XM}} = 5 \cdot 10^{-14}$ m/s/Pa (41). (The effect of different permeabilities is presented in *SI Text*, S2).

**Enzyme kinetics.** We assume segregation of the enzymes in the minor veins and assume that the enzyme-mediated polymerization follows Michaelis-Menten kinetics with a maximal polymerization rate $\phi_{\text{pol}}^{\text{MM}}$ (Fig. 3E-F), and $K_M = 50$ mmol (*Eq. S16*). In Fig. 3A-B, we take $\phi_{\text{pol}}^{\text{MM}} = 900$ nmol/m$^2$/s (the typical export rate in polymer trappers (31)).

**Root kinetics.** We take pressure in the phloem sap at the sinks, $P_S = 0$ MPa; this choice is equivalent to neglecting rate limitations at the unloading step, but variations of the unloading rate can be accounted for by varying the conductance of the transport phloem, $L_P$.

**Plasmodesmatal interface and sugar filtering.** We assume a density of 50 plasmodesmata/μm$^2$ of minor vein wall, which correspond to the upper interval for plasmodesmatal frequency found in the literature (31, 42). We take pore-length of 140 nm, equal to half of the total wall thickness (43), corresponding to the length of the branched side of the plasmodesmata. We treat each plasmodesma as a bundle of $N = 9$ pores (Fig. 2A) (28). For solute hydrodynamic radius, we calculated values from 3D hydrated models of sucrose ($r_\text{suc}$ = 0.42 nm) and stachyose $r_\text{stac}$ (= 0.6 nm) (17). We vary plasmodesmata pore size between $r_\text{pore}$ = 0.6 nm and 0.84 nm (Fig. 3).

**General Modeling hypothesis.** We assume throughout our model that all the supracellular compartments are well mixed and we neglect any gradient of pressure or concentration inside them. The physiological values above and in *Table S1* are of the right order of magnitude and provide a basis for exploring trends, via changes of the non-dimensional flushing number, $f$, reaction rate, $\phi_\text{pol}^\text{MM}$, and confinement parameter $\lambda$. Throughout the study, we vary the flushing number by changing the transport phloem hydraulic permeability, $L_P$ (Table *S1*). At steady state, our model provides fourteen equations (S1-S14) that we solve for the fourteen unknown pressures, concentrations, and fluxes shown in blue and red in Fig. 1*E* (*SI Text*, S1).

## Acknowledgement.

The authors acknowledge fruitful discussions with Kaare Jensen. JC and ADS acknowledge support from the Air Force Office of Scientific Research (FA9550-15-1-0052) and the Camille Dreyfus Teacher-Scholar Awards program. RT acknowledges support from the National Science Foundation- Integrative Organismal Systems (1354718).

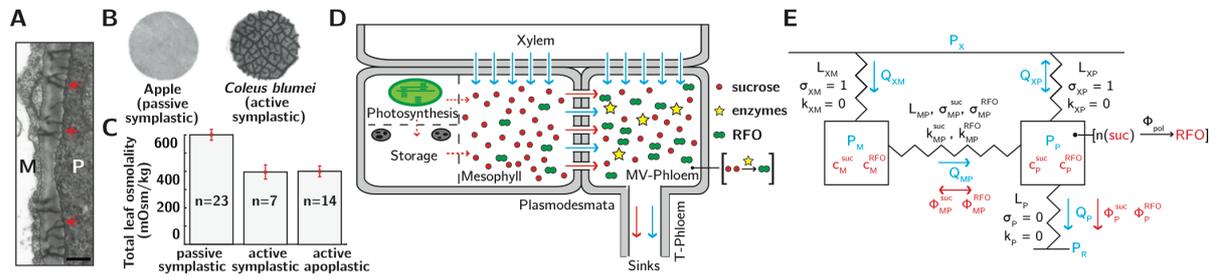

**Figure 1. Overview of phloem loading and global model. (A)** Cross-sectional view of mesophyll/phloem (M/P) interface of a mature *Cucumis melo* leaf, showing plasmodesmata with the secondary branching pattern that is characteristic of active symplastic loaders (arrowheads). Bar = 250 nm. Adapted from (43). **(B)** Autoradiographs of leaf discs from apple (*Malus domestica*), a passive symplastic loader and *Coleus blumei,* an active symplastic loader. Abraded discs were incubated in [$^{14}$C]Suc, washed, freeze dried, and pressed against x-ray film. Minor veins are apparent in *C. blumei,* but not apple discs. Discs are 8 mm diameter. Adapted from (3) and (15) **(C)** Total leaf osmolality in passive and active symplastic and apoplastic loading species. Error bars are standard error; derived from (15). **(D)** Model for water and sugar fluxes in active and passive symplastic loaders. Carbon fixed from $CO_2$ is used to synthesize sucrose (red circles) or is transiently stored as starch. Sucrose passes through plasmodesmata down a concentration gradient from the mesophyll to the phloem. In active symplastic loaders, most of the sucrose entering the phloem is polymerized into RFO (green) by an enzymatic process (yellow stars). Depending on the plasmodesmatal properties, some of the RFO can diffuse back to the mesophyll cells. Sucrose and RFO are exported via bulk flow in the transport phloem. **(E)** Circuit diagram of model in (D). Hydraulic interfaces are characterized by hydraulic permeabilities ($L$ [m s$^{-1}$ Pa$^{-1}$]) and reflection coefficients ($\sigma$ [-]) ($\sigma$=1 for osmotic membranes); the plasmodesmata interface is further characterized by a diffusive mass transfer coefficient ($k$ [m s$^{-1}$]). Volumetric fluxes of water ($Q$ [m s$^{-1}$] – blue arrows) and molar fluxes of solute ($\phi$ [mole m$^{-2}$ s$^{-1}$] – red arrows) pass through the circuit from the xylem at pressure, $P_X$ [Pa] to tissue sinks for the sugars at a pressure, $P_S$ [Pa]. In the MV-phloem, *n* sucrose are polymerized to form one RFO at a rate $\phi_{pol}$ [mole m$^{-2}$ s$^{-1}$]. See *Table S1* for values of all parameters used.

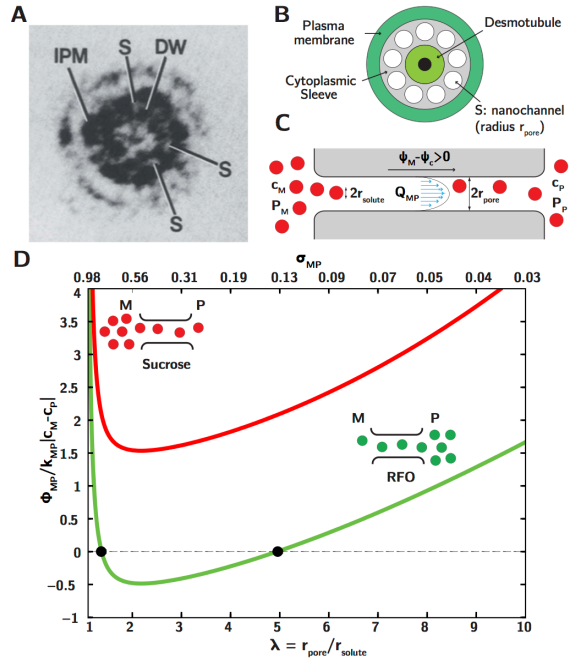

**Figure 2. Plasmodesmata transport.** (**A**) Transmission electron micrograph showing a transverse cross-section of a plasmodesma between phloem parenchyma cells (Fig. 1A presents longitudinal cross-section). Note spaces (S) between particles of the desmotubule wall (DW) and the inner leaflet of the plasma membrane (IPM) (27). (**B**) Schematic representation of longitudinal cross-section of a plasmodesma, showing the desmotubule (a tube of appressed endoplasmic reticulum that extends between the adjacent cells); and the cytoplasmic sleeve between the desmotubule and plasma membrane. Membrane proteins are thought to divide the cytoplasmic sleeve into nanochannels (S) which, though irregular in form, are represented as tubes (inspired by (44)). (**C**) Schematic representation of longitudinal cross-section of a nanochannel. Molecules of hydrodynamic radius $r_{solute}$ are transported by convection ($Q_{MP}$) and diffusion through a nanochannel of radius $r_{pore}$, by diffusion and flow of water created by a water potential difference $\psi_M$-$\psi_P$ between mesophyll (M) and phloem (P). (**D**) Ratio of total and diffusive solute transport in a channel submitted to a water potential difference of 0.1 MPa ($P_M$>$P_P$) as a function of relative pore size, $\lambda = r_{pore}/r_{solute}$ (bottom axis) or equivalent reflection coefficient, $\sigma_{MP}$ (upper axis) (Eq. 8). The gradient of solute is either with (red - sucrose) or against (green - stachyose) the direction of water flow. See *SI Text*, S1 for details on the plasmodesmata transport model.

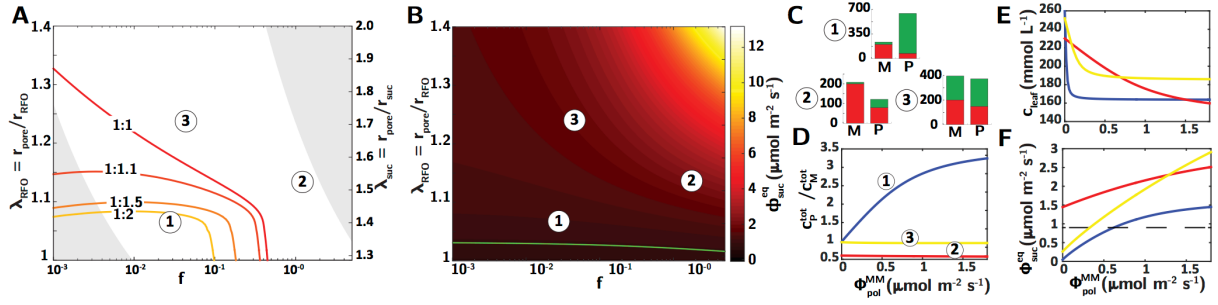

**Figure 3. Gradient inversion and export with hindered transport through plasmodesmata.**
(A) State diagram of gradient inversion as a function of degree of confinement, $\lambda_{RFO}$ and $\lambda_{suc}$ and flushing number, $f$. Isolines show the ratio of the total concentration in the phloem and in the mesophyll, for $r_{stac}/r_{suc} = 1.4$ and hindered plasmodesmatal transport. The red curve (1:1) is the frontier between conditions that provide gradient inversion (minor vein phloem concentration greater than mesophyll concentration, ($c_P>c_M$) and those that do not ($c_P<c_M$). The curves 1:1.1, 1:1.5, and 1:2 correspond to 10, 50, and 100% excess concentration in the phloem. Point 1: $r_{stac}/r_{suc} = 1.13$ and $f = 0.9$. Point 2: $r_{stac}/r_{suc} = 1.23$ and $f = 0.04$. Point 3 : $r_{stac}/r_{suc} = 1.07$ and $f = 0.03$. In constructing this plot, we varied $r_{pore}$ and $L_P$ keeping other parameters fixed (see *Materials and Methods* and *Table S1*). The grey shaded areas represent conditions outside of the estimated physiological range based on $L_p$ (bottom left boundary, $L_p=10^{-16}$ m/s/Pa; top right boundary, $L_p=10^{-12}$ m/s/Pa, Eq. S16). We discuss other scenarios in the SI. (B) Total translocation rate of equivalent sucrose as a function of $\lambda$ and $f$. Black : low translocation rates. Yellow : high translocation rates. The green lines corresponds to a constant export rate of 900 nmol/m2/s, corresponding to typical physiological values (31). (C) Histograms showing sucrose (red) and RFO (green) levels in the mesophyll (M) and Phloem (P) for the conditions of the three points indicated in (A) and (B). (D-F) Plots of ratio of total concentration in the phloem over total concentration in the mesophyll (D), total concentration in the leaf generated for a constant export rate equal to the export rate at zero polymerization (D) and equivalent sucrose flux (E) for the three points in (A) and (B). Blue line, point (1); red line, point (2); yellow line, point (3).

# SUPPLEMENTARY INFORMATION

# Phloem loading through plasmodesmata: a biophysical analysis.

Jean Comtet[†,*], Robert Turgeon[‡,*] and Abraham D. Stroock[†,§,*]

[†]School of Chemical and Biomolecular Engineering,[‡] Section of Plant Biology, and [§]Kavli Institute at Cornell for Nanoscale Science, Cornell University, Ithaca, NY 14853.

*Correspondence: Jean Comtet (jean.comtet@gmail.com), Robert Turgeon (ert2@cornell.edu) and Abraham D. Stroock (ads10@cornell.edu)

*SUPPLEMENTARY TEXT*

**SI Text S1. MATHEMATICAL TREATMENT**

**Governing Equations:** The steady state fluxes of water and solutes in the hydraulic circuit shown in Fig. 1E are governed by the following balance and flux equations. All fluxes and permeabilities are expressed per area of minor veins, i.e. per unit area of the bundle sheath-intermediary cell interface.

*Water Balance Equations for mesophyll and phloem compartments:*

$$Q_{\text{MP}} = Q_{\text{XM}} \qquad \text{(S1)}$$
$$Q_{\text{P}} = Q_{\text{MP}} + Q_{\text{XP}} \qquad \text{(S2)}$$

*Water Flux Equations:*

$$Q_{\text{XP}} = L_{\text{XP}} \Delta \Psi_{\text{XP}} \qquad \text{(S3a)}$$
$$Q_{\text{XM}} = L_{\text{XM}} \Delta \Psi_{\text{XM}} \qquad \text{(S4a)}$$
$$Q_{\text{MP}} = L_{\text{MP}} \Delta \Psi_{\text{MP}} \qquad \text{(S5a)}$$
$$Q_{\text{P}} = L_{\text{P}} \Delta P_{\text{PR}} \qquad \text{(S6a)}$$

In Eqs. S3a-S5a, $\Delta \Psi_{\alpha\beta} = \Psi_\alpha - \Psi_\beta$ [Pa] represent the difference in water potential between compartment $\alpha$ and $\beta$. These driving forces account for both mechanical and osmotic pressure differences with the general form presented below in Eq. S15.

*Solute Flux Equations:*
The fluxes from the mesophyll to the phloem can involve both convection and diffusion through the plasmodesmata:

$$\phi_{\text{MP}}^{\text{suc}} = \phi_{\text{MP}}^{\text{suc}}(\Delta c_{\text{MP}}^{\text{suc}}, \Delta P_{\text{MP}}) \qquad \text{(S7)}$$
$$\phi_{\text{MP}}^{\text{RFO}} = \phi_{\text{MP}}^{\text{RFO}}(\Delta c_{\text{MP}}^{\text{RFO}}, \Delta P_{\text{MP}}) \qquad \text{(S8)}$$

The functions $\Phi_{MP}^{suc}$ and $\Phi_{MP}^{RFO}$ account for both convection and diffusion through the plasmodesmata and are given below in Eq. S17.

The fluxes through the transport phloem are purely convective:

$$\phi_P^{suc} = Q_P c_P^{suc} \quad \text{(S9)}$$

$$\phi_P^{RFO} = Q_P c_P^{RFO} \quad \text{(S10)}$$

**Solute Balance Equations:**

$$c_M^{suc} = \text{fixed} \quad \text{(S11a)}$$

or

$$\phi_{MP}^{syn} = \phi_{MP}^{suc} \quad \text{(S11b)}$$

$$\phi_{MP}^{RFO} = 0 \quad \text{(S12)}$$

$$\phi_{MP}^{suc} = n\phi_{pol} + \phi_P^{suc} \quad \text{(S13)}$$

$$\phi_P^{RFO} = \phi_{pol} \quad \text{(S14)}$$

Eq. S11a represents the case of constant concentration of sucrose in the mesophyll; Eq. S11b represents the case of a fixed synthesis rate, $\phi_{MP}^{syn}$. Eq. S12 states that there is no net creation or export of stachyose out of the mesophyll. In Eq. S13, $n$ is the degree of polymerization of the RFO in the phloem (e.g., $n = 2$ for stachyose). Eq. S13 states that all sucrose entering the phloem from the mesophyll via the plasmodesmatal interface leaves through the transport phloem in the form of sucrose and RFO. Eq. S14 states that RFO formed in the phloem with a rate $\phi_{pol}$ is exported through the phloem.

**Auxiliary Equations:**

*Water potential driving force across osmotic membranes in Eqs. S3-S4:*

$$\Delta\Psi_{\alpha\beta} = \Delta P_{\alpha\beta} - RT\big[\sigma_{\alpha\beta}^{suc}(c_\alpha^{suc} - c_\beta^{suc}) + \sigma_{\alpha\beta}^{RFO}(c_\alpha^{RFO} - c_\beta^{RFO})\big] \quad \text{(S15)}$$

*Oligomerization reaction rate in Eqs. S13-S14:*

$$\phi_{pol} = \frac{\phi_{pol}^{MM} c_P^{suc}}{K_M + c_P^{suc}} \quad \text{(S16)}$$

We assume that the enzyme-mediated polymerization follows Michaelis-Menten kinetics, and neglect intermediary species formed in the process.

*Transport functions for solute transfer between mesophyll and phloem used in Eqs. S7-S8:*

$$\phi_{MP}^i = [1 - \sigma_{MP}^i(\lambda_i)]Q_{MP} \exp\left[c_M^i + \frac{c_M^i - c_P^i}{\exp(Pe_i) - 1}\right] \quad \text{(S17)}$$

In Eqs. S17, $i = \{suc, RFO\}$, the ratio of solute radius to plasmodesmata pore radius is

$$\lambda_i = \frac{r_i}{r_{pore}} \quad \text{(S18)}$$

the Péclet number (convection/diffusion) for solute transport within the pores of the plasmodesmata is

$$\text{Pe}_{\text{MP}}^i = \frac{[1 - \sigma_{\text{MP}}^i(\lambda_i)]Q_{\text{MP}}}{k_{\text{D}}^i} \qquad (S19)$$

the mass transfer coefficient for the species $i$ is

$$k_{\text{MP}}^i = H(\lambda_i)N\rho\frac{\pi r_{\text{pore}}^2 D_i}{l} \qquad (S20)$$

and the reflection coefficient is

$$\sigma_{\text{MP}}^i(\lambda_i) = 1 - W(\lambda_i) \qquad (S21)$$

In Eq. S20, $D_i$ [m$^2$ s$^{-1}$] is the diffusivity of solute $i$. The functions $H(\lambda)$ in Eq. S20 and $W(\lambda)$ in Eq. S21 are the hindrance factors for diffusion and convection transport of the solute from (1) and (2). We used the following equations from (2): for H($\lambda$), we used Eq. 16 for $0 \leq \lambda \leq 0.95$ and Eq. 15 for $\lambda > 0.95$; for $W(\lambda)$ we used Eq. 18. These functions account for purely steric interactions between the solute, solvent, and the wall of a cylindrical pore. We note that convective transport is less hindered relative to diffusive transport (i.e., $1 \geq W(\lambda) > H(\lambda)$), because solute interaction with the pore wall biases the position of the solute toward the center of the pore channel, where the flow speed is maximal.

**Solving Eqs. 1-14 for fluxes, pressures, and concentrations:**

Taking parameters for the hydraulic interfaces as defined in Fig. 1E, the set of equations (S3-S6) can be rewritten as:

$$Q_{\text{XP}} = L_{\text{XP}}(P_{\text{X}} - P_{\text{M}} + RTc_{\text{M}}^{\text{tot}}) \qquad (S3b)$$

$$Q_{\text{XM}} = L_{\text{XM}}(P_{\text{X}} - P_{\text{M}} + RTc_{\text{P}}^{\text{tot}}) \qquad (S4b)$$

$$Q_{\text{MP}} = L_{\text{MP}}[\, P_{\text{M}} - P_{\text{P}} - RT(\sigma_{\text{MP}}^{\text{suc}}(c_{\text{M}}^{\text{suc}} - c_{\text{P}}^{\text{suc}}) + (\sigma_{\text{MP}}^{\text{RFO}}(c_{\text{M}}^{\text{RFO}} - c_{\text{P}}^{\text{RFO}})] \qquad (S5b)$$

$$Q_{\text{P}} = L_{\text{P}}(P_{\text{P}} - P_{\text{R}}) \qquad (S6b)$$

There are fourteen unknowns shown in blue (hydraulic) and red (solute) in Fig. 1E. This system of equations is made non-linear by the advection of solutes down the transport phloem (Eqs. S13 and S14), for the case when open pores are considered, by Eqs. S11 and S12 due to advection-diffusion process through the plasmodesmatal pores (Eq. S17), and by Michaelis-Menten kinetics (S16b).

We proceed to obtain explicit expressions for the water fluxes by solving the linear Eqs. S1-S6 simultaneously, so as to express water fluxes only in term of the concentrations.

$$Q_{\text{XM}} = Q_{\text{MP}} = \frac{1}{\Lambda^2}\{L_{\text{XM}}(L_{\text{XP}}L_{\text{MP}} + L_{\text{MP}}L_{\text{P}})(RTc_{\text{M}}^{\text{tot}} - P_{\text{X}})$$
$$- L_{\text{XM}}L_{\text{XP}}L_{\text{MP}}(RTc_{\text{P}}^{\text{tot}} - P_{\text{X}}) \qquad (S22a)$$
$$- L_{\text{MP}}L_{\text{XM}}(L_{\text{MP}} + L_{\text{MP}})[\sigma_{\text{MP}}^{\text{suc}}RT(c_{\text{M}}^{\text{suc}} - c_{\text{P}}^{\text{suc}})$$
$$+ \sigma_{\text{XM}}^{\text{RFO}}RT(c_{\text{M}}^{\text{RFO}} - c_{\text{P}}^{\text{RFO}})] - L_{\text{XM}}L_{\text{P}}L_{\text{MP}}P_{\text{R}}\}$$

$$Q_{XP} = \frac{1}{\Lambda^2} \{ L_{XP}(L_{XM}L_{MP} + L_{XM}L_P + L_{MP}L_P)(RTc_P^{tot} - P_X)$$
$$- L_{XM}L_{XP}L_{MP}(RTc_M^{tot} - P_X)$$
$$+ L_{XM}L_{XP}L_{MP}[\sigma_{MP}^{suc}RT(c_M^{suc} - c_P^{suc})$$
$$+ \sigma_{XM}^{RFO}RT(c_M^{RFO} - c_P^{RFO})] - L_{XP}L_P(L_{XM} + L_{XM})P_R \}$$
(S22b)

$$Q_P = \frac{1}{\Lambda^2} \{ L_{XM}L_{XP}L_{MP}(RTc_M^{tot} - P_X) + L_{XP}L_P(L_{XM} + L_{MP})(RTc_P^{tot} - P_X)$$
$$- L_{XM}L_{XP}L_{MP}[\sigma_{MP}^{suc}RT(c_M^{suc} - c_P^{suc})$$
$$+ \sigma_{XM}^{RFO}RT(c_M^{RFO} - c_P^{RFO})] - L_P(L_{XP}L_{XM} + L_{XM}L_{MP}$$
$$+ L_{XP}L_{MP})P_R \}$$
(S22c)

where $c_M^{tot} = c_M^{suc} + c_M^{RFO}$, $c_P^{tot} = c_P^{suc} + c_P^{RFO}$, and

$$\Lambda^2 = L_P L_{XP} + L_P L_{XM} + L_P L_{MP} + L_{XP}L_{XM} + L_{XP}L_{MP} \quad (S23)$$

We plug Eqs. S7-S10 and Eqs. S22 into Eqs. S11-S14. For the case of constant concentration of sucrose in the mesophyll (S11a), we are left with 3 non-linear equations to solve numerically in term of the 3 concentrations ($c_M^{RFO}, c_P^{suc}, c_P^{RFO}$). For the case of fixed synthesis rate (S11b), we are left with 4 non-linear equations to solve in term of the 4 concentrations ($c_M^{suc}, c_M^{RFO}, c_P^{suc}, c_P^{RFO}$). We use Matlab (*fmincon*) to find solutions for the solute concentrations. With these values, we can return to Eqs. S22 to find water fluxes, Eqs. S3-S6 to find pressures, and Eqs. S11-S14 to find solute fluxes.

## SI Text S2. AN ALTERNATIVE SCENARIO FOR WATER TRANSPORT

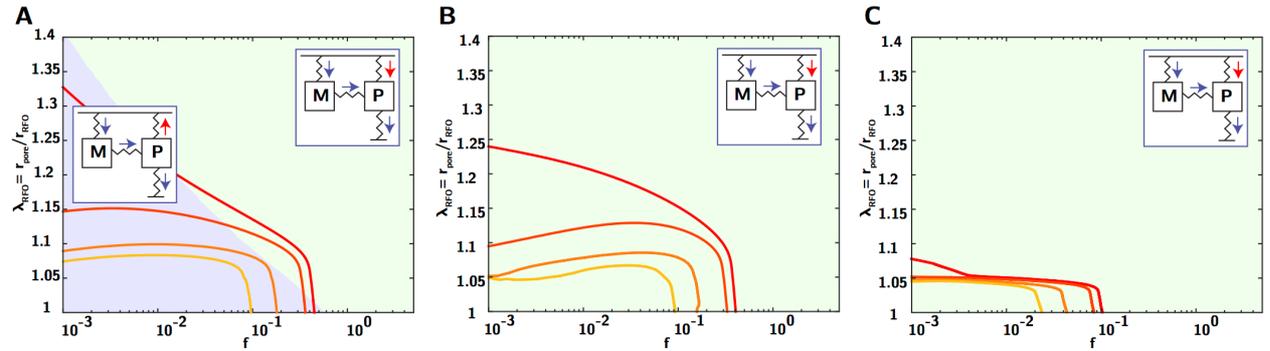

**Figure S1**: Effect of xylem to phloem ($L_{XP}$) and xylem to mesophyll ($L_{XM}$) permeabilities on segregation levels. **(A)** Equal permeabilities. $L_{XP} = L_{XM} = 5 \cdot 10^{-14}$ m/s/Pa, as in Fig. 3A. **(B)** Lower permeability from xylem to phloem. $L_{XM} = 5 \cdot 10^{-14}$ m/s/Pa and $L_{XP} = 5 \cdot 10^{-16}$ m/s/Pa. **(C)** Lower permeability from xylem to mesophyll. $L_{XM} = 5 \cdot 10^{-16}$ m/s/Pa and $L_{XP} = 5 \cdot 10^{-14}$ m/s/Pa. Red to orange contours represent lines of constant levels of gradient inversion of 0%, 10%, 50% and 100% (as in Fig. 3A). Green areas corresponds to zones of the state diagram where water is flowing from xylem to phloem ($Q_{XP} > 0$) (top right inset). Blue areas corresponds to the zones in the state diagram where some water flows from the phloem back into the xylem ($Q_{XP} < 0$, left inset in (A), see text for details).

The flushing number which we introduce in Eq. (4) of the main paper does not capture the effect of different relative coupling between the xylem and the phloem and the xylem and the mesophyll. Different relative values of the permeabilities of the interfaces with the xylem ($L_{XM}$ and $L_{XP}$) impact the path followed by water through the network and can influence the strengths of segregation and gradient inversion observed. We consider three cases:

(1) The case where both membranes have the same permeabilities ($L_{XM} = L_{XP}$) is the one presented in the text (Figs. 4) and shown in Fig. S3A. As we discussed in the main text, segregation (Fig. 3C) and gradient inversion (Fig. 3A) can occur in this case. In Fig. S3A, we show that the region in which gradient inversion occurs (below the red curve) overlaps with a region in which some flow of water actually passes from the phloem into the xylem (blue-shaded zone; $\phi_{XP} < 0$; red arrow inset on left); a steady flow is driven by the distribution of osmolytes around a local circuit from the mesophyll into the phloem and back into the xylem. We note that this circulation actually strengthens the segregation of RFO by increasing the flux through the plasmodesmatal interface and raising the Péclet number for RFO within the pores.

(2) The case where the permeability from the xylem to the mesophyll is larger than the permeability from xylem to phloem ($L_{XM} \gg L_{XP}$) is shown in Fig. S3B. In this situation, $\phi_{XP} > 0$ (green-shaded area) on almost the entire state diagram, and virtually all water export through the transport phloem is flowing through the plasmodesmata ($\phi_{MP} \approx \phi_P$). Importantly for our conclusions in the main text, gradient inversion still occurs in this case, although it requires slightly higher levels of confinement in the plasmodesmatal pores compared to Fig.

S3A (i.e., the isolines of gradient inversion are shifted to lower values of $\lambda_{RFO}$), because flow through the plamodesmata is not as large as in case (1) above.

(3) In the opposite limit where the permeability from the xylem to the mesophyll is smaller than the permeability from xylem to phloem ($L_{XM} \ll L_{XP}$), gradient inversion can still occur, but for even larger levels of confinement in the plasmodesmatal pores (smaller $\lambda_{RFO}$), because the proportion of water flowing through the plasmodesmata is largely reduced compared to the two cases above.

# SUPPLEMENTARY TABLE

## SI Table S1. TABLE OF PARAMETERS

| Notation | Definition | Typical Values |
|---|---|---|
| **Concentrations [mmol]** | | |
| $c_M^{suc}$ | Sucrose concentration in the Mesophylls | 200 mmol |
| $c_M^{RFO}$ | Stachyose concentration in the Mesophylls | - |
| $c_P^{suc}$ | Sucrose concentration in the Minor Vein Phloem | - |
| $c_P^{RFO}$ | Stachyose concentration in the Minor Vein Phloem | - |
| **Permeabilities [m/s/Pa]** | | |
| $L_{XM}$ | Xylem/Mesophylls permeability (3) | $5.10^{-14}$ m/s/Pa |
| $L_{XP}$ | Xylem/Phloem permeability (3) | $5.10^{-14}$ m/s/Pa |
| $L_{MP}$ | Mesophylls/Phloem plasmodesmatal permeability | $10^{-13} - 5.10^{-12}$ m/s/Pa |
| $L_P$ | Transport Phloem equivalent permeability | $10^{-10} - 10^{-16}$ m/s/Pa |
| **Pressures and water Potentials [bar]** | | |
| $P_X$ | Xylem Water pressure or water potential | -1 bar |
| $P_R$ | Root water pressure of water potential | 0 bar |
| $P_M$ | Mesophyll hydrostatic pressure | - |
| $P_P$ | Minor Veins hydrostatic pressure | - |
| **Water Flux [m/s]** | | |
| $Q_{XM}$ | Water flux from Xylem to Mesophylls | - |
| $Q_{XP}$ | Water flux from Xylem to Minor Vein Phloem | - |
| $Q_{MP}$ | Plasmodesmatal Water flux from Mesophylls to Phloem | - |
| $Q_P$ | Water Flux through the transport phloem | - |
| **Sugar Flux through plasmodesmata [mmol/m$^2$/s]** | | |
| $\phi_{MP}^{suc}$ | Sucrose flux through the plasmodesmata | - |
| $\phi_{MP}^{stac}$ | Stachyose flux through the plasmodesmata | - |
| $\phi_{MP}^{syn}$ | Expected synthetic rate in the mesophyll, equal to the flux exported through the phloem at steady-state (4) | 900 nmol/m$^2$/s |
| **Enzyme Kinetics** | | |
| $\phi_{pol}$ | Polymerization rate of sucrose into stachyose [mol/m$^2$/s] | - |
| $\phi_{pol}^{MM}$ | Michaelis-Menten Maximal rate [mol/m$^2$/s] | 900 nmol/m$^2$/s |
| $K_M$ | Michaelis-Menten constant | 50 mmol |
| **Plasmodesmatal Transport Parameters** | | |
| $D_{suc}$ | Cytosolic Sucrose diffusion coefficient [m$^2$/s] (5) | $2.3 \cdot 10^{-10}$ m$^2$/s |
| $D_{stac}$ | Cytosolic Stachyose diffusion coefficient [m$^2$/s] (6) | $1.9 \cdot 10^{-10}$ m$^2$/s |
| $k_D^{suc/stac}$ | Sucrose/Stachyose plasmodesmatal mass transfer coefficient [m/s] | |
| $\sigma_{suc/stac}$ | Sucrose/Stachyose reflection coefficient [-] | 0 - 1 |

| $H_{\text{suc/stac}}$ | Sucrose/Stachyose diffusive hindrance [-] | 0 - 1 |
|---|---|---|
| $W_{\text{suc/stac}}$ | Sucrose/Stachyose convective hindrance $(W \sim 1 - \sigma)$ [-] | 0 - 1 |
| $\rho$ | Plasmodesmatal density [m$^{-2}$] (4, 7) | 50 /μm$^2$ |
| N | Number of pores per plamodesmatas (8) | 9 |
| $r_{\text{pore}}$ | pore radius [m] (4) | 0.7-1.5 nm |
| $l_{\text{pore}}$ | pore length [m] (9) | 140 nm |
| $r_{\text{suc}}$ | sucrose radius [m] (9) | 0.42 nm |
| $r_{\text{stac}}$ | stachyose radius [m] (9) | 0.6 nm |
| $\eta_e$ | Effective phloem sap viscosity including the effects for sieve plates (10) | 5 cPs |
| $\eta_c$ | Typical cytoplasmic viscosity | 2 cPs |
| **Global physiological parameters** | | |
| $v_M$ | Volume fraction of mesophyll is the leaf (13) | 97 % |
| $v_P$ | Volume fraction of phloem in the leaf (13) | 3% |
| $a$ | Sieve tube radius [m] | 5-20 μm |
| $l_{\text{load}}$ | Length of the loading zone (leaf length) [m] | 1-50 cm |
| $h$ | Length of the transport zone (plant height) [m] | 0.1-10 m |

*Hydraulic permeability of plasmodesmatal interface and transport phloem*:

The permeability of the interface between the mesophyll and the phloem has the form:

$$L_{\text{MP}} = N\rho \frac{\pi r_{\text{pore}}^4}{8\eta_c l_{\text{pore}}} \quad \text{(S15)}$$

where *N* is the number of number of effective nanopores per plasmodesma (*N* = 9 in this study), $\rho$ [m$^{-2}$] is the areal density of plasmodesmata, $r_{\text{pore}}$ [m] and $l_{\text{pore}}$ [m] are the effective radius and length of the pores in the plasmodesmata, and $\eta_c$ [kg m$^{-1}$ s$^{-1}$] is the viscosity of the sap. This $L_{\text{MP}}$ varies between 10$^{-14}$ and 5×10$^{-13}$ (m s$^{-1}$ Pa$^{-1}$) for typical values of the parameters characterizing plasmodesmata ($r_{\text{pore}} \in [0.6; 1.5]$ nm)

The hydraulic permeability of the transport phloem has the form:

$$L_{\text{P}} = \frac{a}{l_{\text{load}}} \cdot \frac{a^2}{16\eta_c h} \quad \text{(S16)}$$

where *a* [m] is the radius of the sieve tube, and $l_{\text{load}}$ [m] is the length over which loading occurs (approximately leaf length) (Kåre Hartvig Jensen, Liesche, Bohr, & Schulz, 2012), and *h* [m] is the length of the transport flow. This permeability is expressed for water flow per area of minor vein, leading to an additional geometrical factor $a/l_{\text{load}}$. The range of parameters values are $h \in [0.1; 10]$, $a \in [5; 20]$ μm, and $l_{\text{load}} \in [1, 50]$ cm. The additional hydraulic resistance of the transport phloem due to sieve plates corresponds to approximately half of the total hydraulic resistance, and can be accounted for using an effective viscosity $\eta_e \approx 5$ cPs (10). Taking extreme values in the range above, we obtain transport phloem permeability, $L_P$ in the range of 10$^{-12}$ to 10$^{-16}$ m/s/Pa (grey shaded areas, Fig. 4A-B of the main text). Note that because long transport distances and loading lengths correlate with larger sieve elements (12), we expect $L_P$ to be centered around $10^{-14}$ m/s/Pa.

**Volume Fraction**

Assuming a vein density of 2.45 mm veins per mm$^2$ leaf area, phloem cell cross-sectional area of approximately 250 µ$^2$ (13) and leaf thickness of 200 µm, the volume of minor vein phloem is 3% of that of total leaf tissues.

**SUPPLEMENTARY FIGURES**

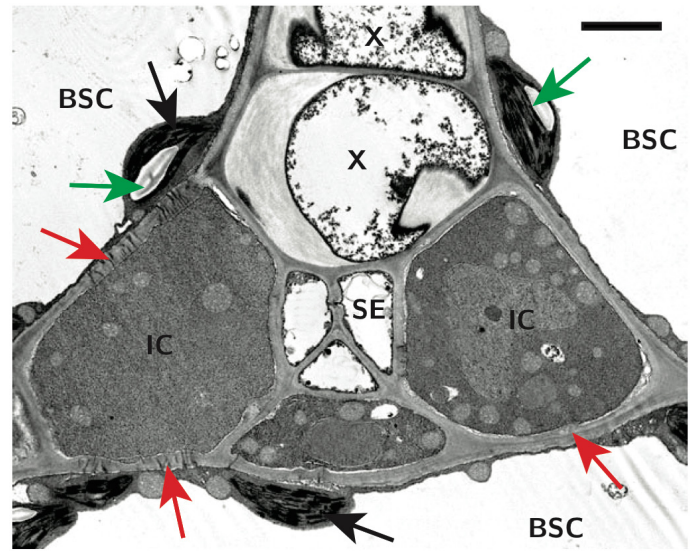

**Figure S2:** Labelled electron micrograph corresponding to the model of Fig. 1D. Transverse section of the minor vein from an active symplastic loader, V. phoeniceum. Intermediary cells (IC) are arranged in two longitudinal files on the abaxial (lower) side of the vein, and each is adjacent to a sieve element (SE). A xylem tracheid (X) is also present. Bundle sheath cells (BSC) are the component of the mesophyll that directly surround the xylem and ICs. The IC:BSC interface, showing numerous plasmodesmata is indicated by red arrows. Chloroplasts (green arrows) and starch (black arrows) are present in BSCs. Scale bar: 1 mm. Adapted fom (14).